\documentclass[
aps,10pt,
prd,
preprintnumbers,amsmath,amssymb,
reprint,
superscriptaddress,
nofootinbib,
a4paper,
twocolumn
]{revtex4-2} 

\usepackage[utf8]{inputenc}
\usepackage[T1]{fontenc}
\usepackage{lmodern}
\usepackage{amsmath,amssymb,amsfonts}
\usepackage{graphicx}
\usepackage{hyperref}
\usepackage{xcolor}
\usepackage{mathtools}
\usepackage{bm}
\usepackage{orcidlink} 

\hypersetup{
	colorlinks=true,
	linkcolor=blue,
	citecolor=blue,
	urlcolor=blue
}

\newcommand\nn{\nonumber\\}
\newcommand\bb{\bibitem}

\newcommand{\bma}{\left(\begin{array}}
	\newcommand{\ema}{\end{array}\right)}
\newcommand{\be}{\begin{equation}}
	\newcommand{\ee}{\end{equation}}
\newcommand{\ben}{\begin{equation*}}
	\newcommand{\een}{\end{equation*}}
\newcommand{\ba}{\begin{eqnarray}}
	\newcommand{\ea}{\end{eqnarray}}
\newcommand{\ban}{\begin{eqnarray*}}
	\newcommand{\ean}{\end{eqnarray*}}
\newcommand{\bs}{\begin{subequations}}
	\newcommand{\es}{\end{subequations}}
\newcommand{\bc}{\begin{center}}
	\newcommand{\ec}{\end{center}}

\def\ds{d_{\rm S}}
\def\dh{d_{\rm H}}
\def\Arg{{\rm Arg}}

\newcommand{\Pl}{{\text{\tiny Pl}}}
\newcommand{\lp}{\ell_\Pl}

\newcommand{\Mpl}{M_\Pl}

\newcommand{\au}[2]{#1.~#2}
\newcommand{\arX}[1]{\href{http://arxiv.org/abs/#1}{{\cob arXiv:#1}}}
\newcommand{\oarX}[1]{\href{http://arxiv.org/abs/#1}{{\cob #1}}}

\newcommand{\book}[5]{\emph{#1} (#2, #3, #4, #5)}
\newcommand{\doin}[6]{\href{https://doi.org/#1}{{\cob {#2 #3} {\bf #4}, #5 (#6)}}}
\newcommand{\doie}[4]{\href{https://doi.org/#1}{{\cob {\bf #2}, #3 (#4)}}}

\newcommand{\doinn}[5]{\href{https://doi.org/#1}{{\cob {#2} {\bf #3}, #4 (#5)}}}
\newcommand{\doij}[5]{\href{https://doi.org/#1}{{\cob {#2} {#3} (#5) #4}}}

\newcommand{\ndoinn}[5]{\href{#1}{{\cob {#2} {\bf #3}, #4 (#5)}}}
\newcommand{\edbook}[6]{\emph{#2}, edited by #1 (#3, #4, #5, #6)}
\newcommand{\edbooks}[5]{\emph{#2}, edited by #1 (#3, #4, #5)}
\newcommand{\procsinm}[5]{in \emph{#1}, edited by #2 (#3, #4, #5)}

\newcommand{\procm}[6]{in \emph{#1}, edited by #2 (#3, #4, #5, #6)}
\newcommand{\tia}[1]{{#1},}

\renewcommand{\geq}{\geqslant}
\newcommand{\Eq}[1]{(\ref{#1})}
\newcommand{\Eqq}[1]{Eq.~(\ref{#1})}
\newcommand{\Eqqs}[1]{Eqs.~(\ref{#1})}
\def\rme{e}
\def\rmd{d}
\def\rmi{i}

\def\Re{\text{Re}}

\def\a{\alpha}

\def\de{\delta}

\def\g{\gamma}
\def\la{\lambda}

\def\Om{\Omega}
\def\om{\omega}
\def\G{\Gamma}

\def\s{\sigma}

\def\B{\Box}

\def\lst{\ell_*}

\def\cF{\mathcal{F}}

\def\cK{\mathcal{K}}
\def\cL{\mathcal{L}}

\def\cO{\mathcal{O}}
\def\cP{\mathcal{P}}

\def\cV{\mathcal{V}}

\def\ups{u}
\def\p{\partial}

\def\cob{\color{blue}}

\begin{document}
	
	\title{Fractal universe and quantum gravity made simple}
	
	\author{Fabio Briscese\,\orcidlink{0000-0002-9519-5896}}
	\affiliation{Dipartimento di Architettura, Universit\`a Roma Tre, Via Aldo Manuzio 68L, 00153 Rome, Italy}
	\affiliation{Istituto Nazionale di Fisica Nucleare, Sezione di Roma 3, Via della Vasca Navale 84, 00146 Rome, Italy}
    \affiliation{Istituto Nazionale di Alta Matematica Francesco Severi, Gruppo Nazionale di Fisica Matematica, Piazzale Aldo Moro 5, 00185 Rome, Italy}

	\author{Gianluca Calcagni\,\orcidlink{0000-0003-2631-4588}}
	\email[Corresponding author: ]{g.calcagni@csic.es}
	\affiliation{Instituto de Estructura de la Materia, CSIC, Serrano 121, 28006 Madrid, Spain}
	
	
\begin{abstract}
Quantum field theory (QFT) on fractal spacetimes is a program aiming at quantizing the gravitational interaction consistently at all energy scales thanks to an intrinsically or dynamically induced multiscale or multifractal-like spacetime geometry that regularizes the infinities of standard QFT. We reach the goal of this program and formulate a field theory of quantum gravity which is shown to be super-renormalizable and unitary at all perturbative orders. Viable and unviable ways to test this proposal through black holes and gravitational waves are discussed.
\end{abstract}

\preprint{\doin{10.1103/kbkv-vs5t}{PHYSICAL REVIEW}{D}{114}{046011}{2026} [\arX{2603.24593}]\hspace{6.5cm} March 25, 2026}
	
\maketitle


\section{Introduction}

The hope to quantize gravity consistently has been fueled by the expectation that quantum mechanics and general relativity should coexist in a unified framework. Proposals in this direction abound \cite{Ori09,Fousp,Calcagni:2013hv,Bam24} and a universal feature of all of them is that, when a spacetime continuum is reached, its spectral dimension $\ds$ changes with the probed scale and becomes 4 only in the infrared (IR) limit \cite{tHooft:1993dmi,Ambjorn:2005db,Lauscher:2005qz,Benedetti:2008gu,Modesto:2008jz,Horava:2009if,Carlip:2009kf,Benedetti:2009ge,Modesto:2009qc,Calcagni:2013eua,Calcagni:2016edi,Carlip:2017eud,Mielczarek:2017cdp,Carlip:2019onx}. This ``dimensional flow'' results in a geometry with characteristics resembling those of self-similar fractal sets and, more generally, of systems with a multiple dependence on some characteristic scales.

Motivated by this theoretical observation, in 2009 a new program was initiated to study the properties and phenomenology of field theories living on multifractal or, more generally, multiscale spacetimes \cite{Calcagni:2009kc,Calcagni:2010bj,Calcagni:2010pa}. This program revived very early attempts to place fields on fractals \cite{Stillinger:1977mt,Svozil:1985ha,Eyink:1989dv,Eyink:1989se} and poses two precise questions: (I) Given that all known proposals of quantum gravity display dimensional flow, is it possible to build a consistent, ultraviolet (UV) complete perturbative quantum field theory (QFT) of gravity (plus matter) featuring this universal characteristic by construction? (II) Do renormalizability and unitarity improve with dimensional flow and, if so, under what conditions?

Dimensional flow as defined above can be realized as an intrinsic property of QFT in a variety of ways \cite{Calcagni:2016azd,Calcagni:2021ipd}. One can modify the integro-differential structure of the action either in the integration \cite{Calcagni:2011kn,Calcagni:2011sz}, or in the integration measure weight \cite{Calcagni:2016xtk,Calcagni:2017via}, or in the derivative operators \cite{Calcagni:2018dhp,El-Nabulsi:2021uwp}, or in all these elements at the same time \cite{Calcagni:2011kn,Calcagni:2011sz,El-Nabulsi:2021uwp,Vacaru:2010wn}. As a rule of the thumb, the less radical the modification of calculus (for instance, only by introducing a nontrivial measure weight in the action and in the ordinary derivatives of the kinetic term \cite{Calcagni:2021ipd,Calcagni:2016xtk,Calcagni:2017via,Che05,Calcagni:2012zj}) the more difficult the embedding of such a structure in QFT \cite{Calcagni:2016azd,Calcagni:2013qqa}, but the easier the extraction of phenomenology and testing for particle physics \cite{Addazi:2018xpt,Calcagni:2015mxx,Calcagni:2015xcf}, black holes \cite{Calcagni:2017ymp,Huang:2020lbv,Pantig:2024fbh}, gravitational waves (GWs) \cite{Yunes:2016jcc,Calcagni:2016zqv,Berti:2018cxi,LIGOScientific:2019fpa,Calcagni:2019kzo,LISACosmologyWorkingGroup:2019mwx,Calcagni:2019ngc,Carson:2019kkh,Calcagni:2020tvw,LIGOScientific:2026fcf}, and cosmology \cite{Sheykhi:2013rla,Shchigolev:2013jq,Calcagni:2013yqa,Rami:2015kha,Calcagni:2016ofu,Das:2018bxc,Maity:2019qbv,Ghaffari:2019qcv,Jawad:2019xdv,Calcagni:2020ads,Garcia-Aspeitia:2022uxz,Kotal:2025hus}. Conversely, fractional calculus is often used without justification from any UV physics, thus making the replacement of ordinary derivatives with fractional ones rather \emph{ad hoc} \cite{Moniz:2020emn,Barrientos:2020kfp,Shchigolev:2021lbm,Jalalzadeh:2021gtq,Jalalzadeh:2022uhl,Rasouli:2022bug,deOliveiraCosta:2023srx,Jalalzadeh:2025uuv}. All of this puts the description of what a realistic fractal quantum universe could look like on shaky ground.

The purpose of this paper is twofold. On one hand, to help all this research converge into a single top-down fractional QFT, which we hereby construct from first principles using knowledge currently scattered in the literature. This QFT has the most desired properties compatible with dimensional flow, in particular, hermiticity, diffeomorphism (diffeo) invariance, renormalizability, and perturbative unitarity. On the other hand, to explore how we can test the theory at present, especially with black holes and GWs. The ``fractal universe'' of the original proposal \cite{Calcagni:2009kc} is thus simplified to just one class of fractional QFTs, of which fractional quantum gravity (FQG) is the main object of interest.


\section{Fractal universe}

The starting point for a quantum fractal universe is the building of a physical field theory on a multiscale geometry. In the following, we focus our attention on Minkowski spacetime because, on one hand, dimensions are always defined ignoring curvature effects and, on the other hand, here we are taking the orthodox perspective of perturbative quantum gravity where all the properties of the theory are studied in a local inertial frame. Consider a continuous spacetime with $D$ topological dimensions, metric $g_{\mu\nu}$, signature $(-,+,\cdots,+)$, and a generic model action with Minkowski metric $g_{\mu\nu}=\eta_{\mu\nu}$, exotic measure weight $v(x)$, and kinetic operator $\cK(\p_\mu)$:
\be\label{actmodel}
S=\int\rmd^Dx\,v(x)\left[\frac12\phi\,\cK(\p_\mu)\,\phi+\cL_{\rm int}\right],
\ee
where $\phi$ is a real scalar field, $\mu=0,1,\dots,D-1$, and $\cL_{\rm int}$ collects interactions. A geometry can be multiscale if at least one of the two dimensions it is mainly characterized by, the Hausdorff and the spectral dimension, is scale dependent. The Hausdorff dimension $\dh$ is the scaling of a ball hypervolume $\cV(\ell)\sim\int_{\rm ball}\rmd^Dx\,v(x)$ with its linear size $\ell$: $\dh\coloneqq \rmd\ln\cV(\ell)/\rmd\ln\ell=D-[v]$. 
For a geometry with space and time, the spacetime Hausdorff dimension is calculated in Euclidean signature or, alternatively, one can take the time and spatial dimensions separately. The same applies for the Hausdorff dimension $\dh^k$ of momentum space. For multiscale spacetimes with a well-defined IR limit, there are universal factorizable, parametric measures $\rmd^Dx\,v(x)$ and $\rmd^Dk\,\tilde v(k)$ that generalize fractional integrals and entail a certain number of fundamental scales \cite{Calcagni:2016xtk}. These measures break ordinary Lorentz and diffeo symmetries, which can be replaced with generalized ``fractional'' versions by keeping ordinary derivatives with appropriate weights (the so-called $q$-derivatives) in the action \cite{Calcagni:2016azd}. However, the resulting quantum theory does not have improved renormalizability \cite{Calcagni:2021ipd}. Therefore, we are left with the other, complementary possibility of fixing the measures as standard ($\dh=D=\dh^k$) and realizing dimensional flow through the spectral dimension $\ds$. For an ordinary momentum space, $\ds$ is roughly the inverse of the asymptotic scaling of the dispersion relation $\tilde\cK(k)\sim k^{2D/\ds}$, where $\tilde\cK(k)$ is the Fourier transform of the kinetic term. More precisely, $\ds\coloneqq -\rmd\ln\cP(\ell)/\rmd\ln\ell$, where $\cP(\ell)\coloneqq\cP_0 \int\rmd^Dk\,\exp[-\ell^2\tilde\cK(k)]$ is called return probability and the prefactor $\cP_0$ is $\ell$-independent.\footnote{Usually, $\cP$ is defined as the trace over spacetime points $\cP(\ell)\propto \int\rmd^Dx\, \cP(x,x;\ell)$ of the solution of the diffusion equation $[\p_{\ell^2}-\cK(\B_x)]\cP(x,x';\ell)=0$ of a pointwise probe with kinetic term $\cK$ in Euclidean signature and in a local inertial frame, where $\cP(x,x';0)=\de^D(x-x')$. On smooth spacetimes, the definition given in the main text coincides with this one (see, e.g., \cite{Calcagni:2014wba} or question 01 in Ref.~\cite{Calcagni:2016azd}) and is operationally simpler. A reinterpretation of $\cP(\ell)$ as the self-energy at a given resolution $\ell^{-1}$ is also possible \cite{Calcagni:2014wba}.} For example, let $\B=\p_\mu \p^\mu$ be the d'Alembertian [in momentum space, $\B\to -k^ 2=(k^0)^2-|\bm{k}|^2$]. For $\cK=(-\B)^\g$, one has $\tilde\cK(k)=(k^2)^\g$ and $\ds=D/\g$. 

Having fixed the measure of position and momentum space, we determine the kinetic term $\cK$. We must ensure the action to have enough symmetries both to allow for a manageable QFT and to preserve the powerful diffeo invariance of general relativity. Given that the Lebesgue measure is trivial and that geometry is solely specified by the metric, diffeo invariance is equivalent to covariance, hence $\cK=\cK(\B)$ in \Eqq{actmodel}. This marks a shift of interest from coordinate-dependent modifications of the integro-differential structure of the theory (fractional calculus \cite{MR,SKM}) to background-independent covariant ones (fractional operators \cite{Bal60,Kom66,Seeley:1967ea,Hor68}). Moreover, the classical limit of the theory is well defined only if the action \Eq{actmodel} is real, hence only if $\cK^\dagger=\cK$. This precludes non-Hermitian operators such as $\cK=(-\B)^\g$ when $\g$ is noninteger. 

The last step is to encode multiscaling in $\cK$ while respecting all the above constraints. The most rigid and regular multiscale spacetime is self-similar like in deterministic fractals \cite{Calcagni:2016azd,Calcagni:2017via,fge2,NLM,Calcagni:2011nc}, i.e., sets invariant under discrete similarity transformations $x^\mu\to\la_\Om x^\mu$, where $\la_\Om\equiv\exp(-\pi/\Om)$ and $\Om>0$ are fixed constants. This corresponds to a kinetic term with discrete scale invariance (DSI), for example, of the self-adjoint form
\be\label{Klog}
\cK(\B)=\cF(\B)\sum_{n=0}^N\cos[n\Omega\ln(\lst^4\B^2)+\theta_n]\,,
\ee
where $\cF=\cF^\dagger=\B+\dots$ is self-adjoint, $\theta_n$ is a phase parameter, and $\lst$ is a fundamental length scale. Self-adjointness is required in order to have a well-defined classical limit \cite{Calcagni:2025wnn}, so that the ellipsis in $\cF$ can be either a function of generic integer powers of $\B$ or of arbitrary powers of $\B^2$: the latter is precisely the choice of fractional QFT. The function in the sum in \Eqq{Klog} is therefore self-adjoint, too, and is determined by allowing DSI. Under a transformation $x^\mu\to\la_\Om x^\mu$, one has $\ln\B^2\to \ln(\la_\Om^{-2}\B^2)=2\pi/\Om+\ln\B^2$ and the sum is DSInvariant. The whole operator $\cK$ is not invariant because $\cF$ is made of additive factors and has therefore no DSI, but this symmetry, if present, does not need to be exact at all scales and can be confined to the UV \cite{Calcagni:2016azd,Calcagni:2017via,Calcagni:2011nc}. Call $\ell_\Om\equiv\lst\la_\Om<\lst$, where the inequality is due to the fact that $0<\la_\Om<1$. For natural $O(1)$ values $\Om\lesssim 1.4$, this scale is ultramicroscopic, $\ell_\Om\ll \lst$. Self-similarity becomes manifest at observation length scales $\ell\sim \ell_\Om$. At larger but still small scales $\ell_\Om\ll \ell\lesssim\lst$, the log-oscillations in \Eqq{Klog} are smeared, and only the zero mode $\cK(\B)\simeq\cF(\B)$ survives, with its characteristic UV/IR divide determined by $\lst$. At large scales $\ell\gg\lst$, one recovers the standard kinetic operator $\cK(\B)\simeq\cF(\B)\simeq\B$. 

This geometry is invariant under Lorentz transformations $x^\mu=\Lambda_\nu^\mu x^\nu$ in any local inertial frame. Since these are continuous affine transformations, we have a self-affine geometry from the deep (but not ultra) UV to the IR, albeit in the trivial general-relativistic sense.

In this limit, we realize a basic multiscaling corresponding to the UV/IR divide normally found in quantum gravity. We are thus ready to fix $\cF(\B)$. A single fundamental length scale $\lst$ 
 is enough to make the correlation functions $\langle\cO(x)\cO(x')\rangle$ of observables $\cO$ change with the probed energy scale. This type of long-range scaling is typical of random multifractals where the scaling factors of their self-similar transformations have been randomized at each iteration (the above-mentioned smearing limit) \cite{Calcagni:2016azd,NLM,Calcagni:2011nc}. Calling $\g$ the anomalous scaling of the kinetic term in the UV, we are looking for something like $\cF(\B)\sim \B +\lst^{-2}(\lst^2\B)^\g$ but with self-adjointness. $\lst$ marks the transition between a dynamics with second-order derivatives and one of fractional order. While there are infinitely many ways to construct such an operator, a central lesson from multiscale complex systems (ranging from the renormalization group and effective field theory near fixed points \cite{Wilson:1971bg,Fisher:1974zz,Goldenfeld:1992qy} to turbulent, percolating, and critical phenomena \cite{Frisch:1995zz,Barenblatt:1996bg}, to the diffusion techniques on multifractional spacetimes \cite{Calcagni:2012rm}) is that multiscale physics is primarily governed by the asymptotic behavior of correlation functions and is robust against changes of the short-range details of transient regimes. In essence, at the scales $\gg\lst$ where observations are available, it does not matter how the IR term $\B$ and the UV term $\B^\g$ are combined together. We give a proof of this elsewhere \cite{Calcagni:2026sfx} and show here that, for a field with mass $m$, a toy model \Eq{actmodel} with $v=1$ capturing the main symmetry and renormalizability properties of scalar, gauge, and gravitational QFTs has
\be\label{Fdef}
\cF(\B)=F(\lst^2\B-\lst^2m^2)\,,\quad F(z)=z\!\left[1+ (z^2)^\frac{\om}{2}\right]^\ups,
\ee
where $\om\in \mathbb{R}^+$ is noninteger and $\ups\in\mathbb{N}^+$. The operator $F$ is of order $\g\equiv \ups\,\om+1$ and is a generalization of the fractional d'Alembertian $(-\B)^\g$ with noninteger $\g$ \cite{BGG,Mar91,Gia91,BGO,BG,doA92,Barci:1995ad,BBOR1,Barci:1996ny,BBOR2}. It is the minimal self-adjoint realization interpolating between the IR behavior $F\simeq z$ and the UV scaling $F\sim z^{\g}$ and covers both the minimal case of a monomial ``$\B+\B^\g$'' ($\ups=1$) and a more general polynomial of fractional powers of $\Box$ with $\g$ being the highest exponent. Therefore, \Eqq{Fdef} represents a universality class in the sense of the above-mentioned complex systems, i.e., a parametrization of infinitely many operators with the same UV scaling and obeying the logical steps described above. Despite the linear reasoning we offered here to reach \Eqq{Fdef}, in practice it took some time to converge to this point, passing through Euclidean \cite{Trinchero:2012zn,Trinchero:2017weu,Trinchero:2018gwe} and non-Hermitian fractional QFT \cite{Calcagni:2021ljs,Calcagni:2021aap,Calcagni:2022shb}, up to a recent purely fractional self-adjoint model \cite{Calcagni:2025wnn}.

\begin{figure}[ht]
	\bc
	\includegraphics[width=6.5cm]{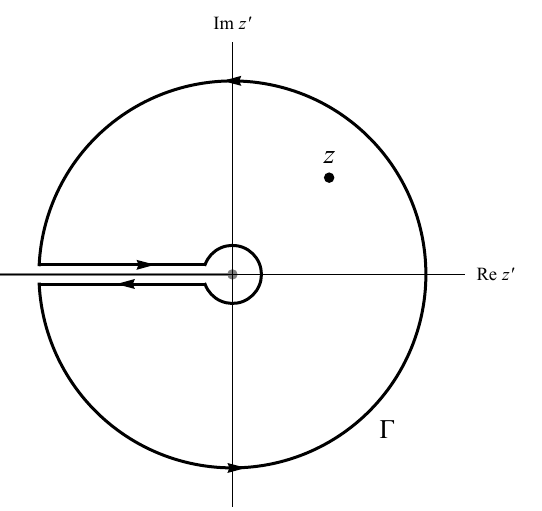}
	\ec
	\caption{\label{fig1} Contour in the Cauchy representation \Eq{d_spectral}.}
\end{figure}

The fractional form factor $F$ admits different representations \cite{Calcagni:2025wnn}. It must fulfill the requirement that the action be real valued both in Euclidean and in Minkowski signature, which implies that $F(z)$ must be defined piecewise. Moreover, the resulting propagator should include the standard IR mode of a massive real particle. In what follows, we derive a suitable $F(z)$. We start defining $F(z)$ on the half-plane corresponding to $\Re\,z>0$ as
\ben
F(z) =f(z)\coloneqq z\left(1+ z^{\om}\right)^\ups\,, \qquad \Re\,z>0\,, \label{definition_F}
\een
with $z^\om$ in the principal branch, so that $z^\om= |z| \, \rme^{\rmi \varphi \om} \in \mathbb{R}^+$ for $z\in \mathbb{R}^+$, where $\varphi = \Arg\,z\in [-\pi,\pi[$. If one defined $F(z)$ for $\Re\,z<0$ by analytic continuation of $f(z)$ over the $z$-plane, then one would obtain a multivalued kinetic term and propagator, which would give a complex-valued action, since $f(z)$ is complex for real negative $z$. Therefore, we need a different route to define $F(z)$ on the whole complex plane. In order to proceed, let us first consider the Green's function corresponding to the choice $F(z)=f(z)$ valid for $\Re\,z>0$, namely,
\be\label{definition_g}
g(z) =-\frac{1}{f(z)}= -\frac{1}{z} \frac{1}{\left(1+z^\om\right)^\ups}\,.
\ee
Our strategy is to find a suitable integral expression for $g(z)$ in the half-plane $\Re\,{z}>0$, and then to extend it to the $\Re\,z<0$ region with a symmetry argument. One might be tempted to extend the propagator as $g(z)\to g(-z)$ in the $\Re{z}<0$ half-plane, corresponding to the assignment of each of the branches of $z\to\sqrt{z^2} = \pm z$ to one half of the complex plane \cite{Calcagni:2025wnn}. However, applying this symmetry to the whole function $g(z)$ would also imply an unphysical change of the sign of the propagator of the massive mode. To avoid that, we separate the $1/z$ contribution in \Eqq{definition_g} as
\be\label{definition_g2}
g(z) =-\frac{1}{z}- \frac{1-\left(1+z^\om\right)^\ups}{z\,\left(1+z^\om\right)^\ups}\,\eqqcolon -\frac{1}{z} + h_\om(z)\,,
\ee
and then work out a suitable integral expression for $h_\om(z)$. Since we are in the half-plane $\Re\,z>0$, taking the principal branch of the square root, we can replace $z\to\sqrt{z^2}$ in $h_\om$ to get
\be\label{definition_hom}
\hspace{-.3cm}h_{\om}(z)=-\frac{1-\left[1+ (z^2)^{\frac\om2}\right]^\ups}{\sqrt{z^2}\left[1+ (z^2)^{\frac\om2}\right]^\ups} \eqqcolon d_{\frac{\om}{2}}(z^2), \quad \Re\, z>0.
\ee
[Note that the more general replacement $z\rightarrow z^{1/N}$ would lead to a different definition $h_\om(z)\eqqcolon d_{\om/N}(z^N)$, which would not correspond to a self-adjoint theory for $N$ odd.] By means of the Cauchy theorem applied to the contour shown in Fig.~\ref{fig1}, where the small and big circles have infinitesimal and infinite radius, respectively, one has
\be\label{d_spectral}
d_\a(z)=\frac{1-\left(1+ z^\a\right)^\ups}{\sqrt{z}\,\left(1+ z^\a\right)^\ups}=\frac{1}{2\pi\rmi}\oint_{\G}\rmd z'\,\frac{ h_\a(z')}{z'-z}\,.
\ee

In order to evaluate this integral, we have to determine the singularities of $d_\alpha(z)$. Equation (\ref{d_spectral}) implies that $d_\alpha(z)$  has a branch point at $z=0$ corresponding to the branch point of $z^\alpha$, and poles at the solutions of $1+z^\a=0$. No such solution appears in the principal branch of $z^\a$ if $\a \in \mathbb{Q}^+$ and $\a<1$. In fact, for $\a$ rational, one can write $\a = p/q$ irreducible with $p,q \in \mathbb{N}$, so that  such roots are  
\be\label{roots}
z_n= \rme^{\rmi \pi (2n+1)\frac{q}{p}}\,,\qquad \quad n=0,1,\ldots,p-1 \,.
\ee
All these roots are outside the interval $\Arg\,z\in [-\pi,\pi[$ if $\pi q/p > \pi$, that is, $\a = p/q <1$. Hereafter, we assume $\a<1$, so that the contour in Fig.~\ref{fig1} does not enclose any singularity. Moreover, since $\alpha = \omega/2$ in \Eq{definition_g2} and \Eq{definition_hom}, this implies that $g(z)$ has a branch point at $z=0$ and no poles in the half-plane $\varphi = \Arg\,z\in [-\pi/2,\pi/2[$ in the principal branch, for $\om<2$. Also, we choose $\ups> 1/\om$ such that $\g=\ups\,\om+1> 2$ in order to include the case of renormalizable FQG (see below).

The contributions  of the infinite and infinitesimal circles in Fig.~\ref{fig1} to the Cauchy integral in \Eqq{d_spectral} are zero, and one is left with the contributions of the branch cut
\ben
d_\a(z)=\frac{1}{2\pi}\!\!\int_0^{+\infty}\!\!\tfrac{\rmd s}{s+z}\tfrac{1}{\sqrt{s}} \left[2-\tfrac{1}{\left(1+\rme^{-\rmi\pi\a} s^\a\right)^\ups}\!-\!\tfrac{1}{\left(1+ \rme^{\rmi\pi\a} s^\a\right)^\ups}\right]\!,
\een
so that, after insertion of this into \Eqqs{definition_g2} and \Eq{definition_hom} with $\a=\om/2$ and a parameter redefinition $s\rightarrow s^2$, we get
\ba
\hspace{-.6cm}g(z)\!&=&\!-\frac{1}{z}+\int_0^{+\infty}\rmd s\,\frac{\rho_{\rm}(s)}{s^2+z^2}\,,\qquad \Re\,z>0\,,\label{definition_g3}\\
\hspace{-.6cm}\rho(s)\! &=& \!\frac{1}{\pi} \left[2-\frac{1}{\left(1+ \rme^{-\rmi\pi\frac{\om}{2}} s^\om\right)^\ups}-\frac{1}{\left(1+ \rme^{\rmi\pi\frac{\om}{2}} s^\om\right)^\ups}\right]\!.\label{rho}
\ea
Incidentally, note that $\rho(s)\geq 0$ for $s\in \mathbb{R}^+_0$ if $0<\om<1$. 

Although $g(z)$ is complex valued for $z \in \mathbb{R}^-$, we can promote \Eqq{definition_g3} to the physical Green's function:
\ba
\hspace{-.7cm}G(z) &\coloneqq& -\frac{1}{z} +\Theta(\Re\,z)\, h_\om(z)+\Theta(-\Re\,z)\, h_\om(-z)\nn
&=&-\frac{1}{z}+ d_{\frac{\om}{2}}(z^2)= -\frac{1}{z}+\int_0^{+\infty}\rmd s\,\frac{\rho_{\rm}(s)}{s^2+z^2}\,.\label{definition_g4}
\ea
This function is defined piecewise, being it discontinuous on the whole imaginary axis of the complex $z$-plane, while it is analytic elsewhere. 


\section{Quantum gravity} 

The above toy model is the basis for a theory of quantum gravity. Let $h_{\mu\nu}=g_{\mu\nu}-\eta_{\mu\nu}=h_+ e_{\mu\nu}^++h_\times e_{\mu\nu}^\times$, where $h_{+,\times}$ are the usual polarization modes of the graviton. Above, we presented the general construction of a covariant fractional QFT with scale-dependent spectral dimension, where fields have a kinetic term with scaling $\cF(\B)\sim \B+\B^\g$. This paradigm should be applied to all the sectors of the theory (scalar, gauge, gravitational, and so on), but in this paper our main interest is in gravity. In particular, the linearized equation of motion for the graviton with universally generic UV anomalous scaling $\sim \B^\g$ has the kinetic term \Eq{Fdef}, $\cF(\B)\,h_{\mu\nu}+\dots=0$, from which one can reconstruct the classical gravitational action for a generic form factor $\cF$ \cite{Calcagni:2021aap}. In $D=4$ dimensions, in our case we obtain
\be\label{fqg}
S=\frac{\Mpl^2}{2}\int\rmd^4x\,\sqrt{|g|}\left[R+G_{\mu\nu}\cF_2(\B) R^{\mu\nu}\right],
\ee
where $\Mpl$ is the reduced Planck mass, $G_{\mu\nu}=R_{\mu\nu}-(1/2)g_{\mu\nu}R$ is the Einstein tensor,
\be\label{fofa}
\cF_2(\B) = \frac{\lst^{-2}\cF(\B)-\B}{\B^2} = \frac{\left[1+ (\lst^4\B^2)^\frac{\om}{2}\right]^\ups-1}{\B}\,,
\ee
and the curvature terms are built from the metric $g_{\mu\nu}$ as usual. Nonlocal $R^2$ and $R_{\mu\nu\s\tau}^2$ terms are possible but unnecessary for renormalizability and unitarity, as shown in the non-Hermitian case \cite{Calcagni:2021aap,Calcagni:2022shb}. A path-integral quantization exists for generic nonlocal theories including FQG \cite{Calcagni:2024xku}, but in the following we focus on the UV properties of the theory \Eq{fqg}. 

Proof of renormalization takes two steps. First, one shows that the Bogoliubov--Parasiuk--Hepp--Zimmermann renormalization scheme \cite{Bogoliubov:1957gp,Hepp:1966eg,Zimmermann:1969jj,Lowenstein:1975ug,Piguet:1995er} holds for fractional QFT \cite{Calcagni:2022shb}. Therefore, all counterterms are local and it is sufficient to show power-counting renormalizability to conclude that the theory is renormalizable. Power counting is the second step. Each internal line scales as $\tilde G(-k^2)\sim (k^2)^{-\g}$ in the UV in momentum space. For gravity, the interaction term in the toy model \Eq{actmodel} is $\cL_{\rm int}\propto-\phi^2\cF(\B)\phi^2$ \cite{Calcagni:2022shb}, so that vertices carry momentum powers ${\rm V}(k)\sim (k^2)^\g$. The integral associated with a generic Feynman diagram with $L$ loops, $N_V$ vertices, and $N_I$ internal lines scales as $(\rmd^4 k)^L [{\rm V}(k)]^{N_V}[\tilde G(-k^2)]^{-N_I}\sim k^{2[L+\g (N_V-N_I)]}$. Using the topological relations $N_V-N_I=1-L$ and $4N_V=2N_I+N_E$, where $N_E$ is the number of external legs, the worst possible superficial degree of divergence $\Delta_{\rm div}$ is for $N_E=2$ and does not exceed 
\ben
\Delta_{\rm div}=2[2-(L-1)(\g-2)]\,.
\een
Divergences appear at $L$ loops if $\Delta_{\rm div}\geq 0$. Therefore, when $\g>2$ the theory is super-renormalizable and, for $\g>4$, we have $\Delta_{\rm div}<4(2-L)$ and the theory is one-loop super-renormalizable (divergences only for $L=1$). 

Unitarity follows from \Eqq{definition_g4} and the fact that the propagator has a continuum of modes with a quartic Green's function $(s^2+\lst^4k^4)^{-1}$. These are pairs of ``$\rmi$-particles'' \cite{Baulieu:2009ha} with complex-conjugate masses, hence they cannot appear in the physical spectrum and must remain off shell and purely virtual at all perturbative orders. In the literature of Yang--Mills models and chromodynamics \cite{Nakanishi:1975aq,Dudal:2005na,Dudal:2007cw,Dudal:2008sp,Baulieu:2009ha,Dudal:2011gd,Capri:2012hh} and of Lee--Wick theories \cite{Modesto:2015ozb,Modesto:2016ofr,Mannheim:2018ljq,Mannheim:2020ryw}, off-shellness is argued to be due to confinement. However, this is not enough because complex-conjugate pairs violate Lorentz invariance \cite{Anselmi:2025uzj,Nakanishi:1971jj}, unless one defines scattering amplitudes with the fakeon prescription \cite{Anselmi:2017yux,Anselmi:2017lia,Anselmi:2018bra,Anselmi:2019rxg,Anselmi:2021hab,Anselmi:2022toe,Anselmi:2025uzj,Anselmi:2025uda}. The latter can be applied to fractional QFT since amplitudes are just convolutions of various propagators $\sim (s^2+\lst^4k^4)^{-1}$, which admit a clean fakeon projection order by order \cite{Anselmi:2025uzj}. Thus, the only asymptotic states in the physical spectrum of the theory are the standard graviton modes $h_{+,\times}$ with propagator $\sim k^{-2}$. FQG is therefore unitary regardless of the sign of $\rho(s)$.



\section{Black holes and gravitational waves} 

We conclude with some applications to compact objects. 
In general, we have a difficult time constraining signatures of FQG with GW observations. For instance, in FQG the graviton has zero mass and the standard dispersion relation $E^2=|\bm{k}|^2$, so that we cannot use any of the constraints on the mass of the graviton and on Lorentz-violating dispersion relations $E^2=|\bm{k}|^2(1+a|\bm{k}|^{1-\a})$ successfully applied to other multiscale models \cite{Yunes:2016jcc,Calcagni:2016zqv,Berti:2018cxi,LIGOScientific:2019fpa,Carson:2019kkh,LIGOScientific:2026fcf}. As another example, bright standard sirens such as GW170817 are sources both of GWs and photons \cite{Tamanini:2016zlh}. The strain measured in a GW interferometer from such an event is given by the ratio between the luminosity distance $d_{\textsc{l}}^{\textsc{gw}}$ measured via GWs and the luminosity distance $d_{\textsc{l}}^{\textsc{em}}$ of its electromagnetic counterpart, which is 1 in Einstein gravity and $d_{\textsc{l}}^{\textsc{gw}}/d_{\textsc{l}}^{\textsc{em}}=1+O(1)(d_{\textsc{l}}^{\textsc{em}}/\lst)^{\G_{\textsc{uv}}-1}$ in multiscale models \cite{Calcagni:2019kzo,Calcagni:2019ngc}, where $\G_{\textsc{uv}}$ is the UV dimensionality of $h_{\mu\nu}$ and of the graviton modes. In FQG,
\be\label{Guv}
\G_{\textsc{uv}}\coloneqq [h_{+,\times}]_{\textsc{uv}}=\frac{D-[\cF]}{2}\simeq 2-\frac{4}{\ds^{\textsc{uv}}}=2-\g<0\,,
\ee
with $\ds^{\textsc{uv}} = 4/\g<2$, and we cannot place any significant constraint on $\lst$ from these kinds of observations, which would require $\G_{\textsc{uv}}\gtrsim 1$ \cite{Calcagni:2019kzo,Calcagni:2019ngc}. For the same reason, in the absence of a screening mechanism, Solar System tests of Newton's law are insensitive to fractional modifications. Newton's potential $\Phi\propto h_{00}$ has been calculated for a generic double-power-law dispersion relation in \cite{Calcagni:2019ngc}. Applying that to FQG (renormalizable if $\g>2$), we get
\be\label{newt}
\Phi(r)\stackrel{\text{\textsc{ir}}}{\simeq} r^{-1}\,,\qquad \Phi(r)\stackrel{\text{\textsc{uv}}}{\sim} r^{-\G_{\textsc{uv}}}= r^{\g-2}\,,
\ee
and the ultrashort scales $r\ll \lst$ in \Eqq{newt} are clearly unreachable at present if $\lst=O(\lp)$.

Black-hole singularities are not avoided in multiscale models with ordinary derivatives and nontrivial measure weights \cite{Calcagni:2017ymp,El-Nabulsi:2021uwp}, nor in gravity models with fractional derivatives \cite{El-Nabulsi:2021uwp,Vacaru:2010wn}. At first, the same would seem to hold in FQG. Indeed, the action \Eq{fqg} admits exact Ricci-flat solutions ($R_{\mu\nu}=0$), including Schwarzschild.
One can include a $R_{\mu\nu\s\tau}^2$ term in \Eqq{fqg} and thus forbid Ricci-flat solutions, but this is not ideal for two reasons: it adds an operator unnecessary at the quantum level and it bars the way to exact black-hole solutions, as is known in other nonlocal theories \cite{Cornell:2017irh,Buoninfante:2018xiw,Koshelev:2018hpt,Buoninfante:2018stt}. Singularity resolution without an $R_{\mu\nu\s\tau}^2$ term can still be achieved at the quantum level by enforcing Weyl symmetry, i.e., when the theory is not only super-renormalizable but also finite (see Ref.~\cite{Modesto:2022asj} for an explanation of the mechanism). This extension is possible if one adds certain $O(R^4)$ nonlocal operators that make all beta functions vanish \cite{Calcagni:2022shb}.

However, and quite remarkably, \Eqq{newt} suggests that there might exist non-Ricci-flat spherically symmetric solutions given by a modified metric with 
$g_{00}=1-(r/r_{\rm s})^{-1}\to 1-(r/r_{\rm s})^{\g-2}$ at small scales ($r_{\rm s}=2GM$ is the Schwarzschild radius), which is finite in the limit $r\to 0$.
This reopens the case for regular black holes, at least of microscopic size $r_{\rm s}=O(\lst)$ (more precisely, $r_{\rm s}<2\lst$ \cite{Buoninfante:2018rlq}), which are expected to be a minimal outcome of any candidate quantum gravity. The shadow and waveform of these objects might deviate from the ones in Einstein gravity and will deserve further scrutiny.
	

\section*{Acknowledgments} 

The authors thank D.~Anselmi and L.~Rachwa\l\ for useful discussions and are supported (G.C.\ as PI) by Grant No.\ PID2023-149018NB-C41 funded by the Spanish Ministry of Science, Innovation and Universities MCIN/AEI/10.13039/501100011\-033. The work of G.C.\ was made possible also through the support of the WOST, \href{https://withoutspacetime.org}{WithOut SpaceTime project}, supported by Grant ID 63683 from the John Templeton Foundation. The opinions expressed in this work are those of the authors and do not necessarily reflect the views of the John Templeton Foundation.

\section*{Data availability} 

There are no publicly available research data or software supporting this manuscript. Requests for further information or data should be sent to the authors.


\end{document}